\documentclass{aastex}
\usepackage{emulateapj5}
\usepackage{apjfonts}
\input psfig.tex
 
\def\aj{{AJ}}

\def\apj{{ApJ}}
\def\apjs{{ApJS}}

\def\baas{{BAAS}}

\def\deg{$^{\circ}$}

\def\dsec{{\rlap.}$^{\prime\prime}$}

\def\gax{{$\mathrel{\hbox{\rlap{\hbox{\lower4pt\hbox{$\sim$}}}\hbox{$>$}}}$}}

\def\lax{{$\mathrel{\hbox{\rlap{\hbox{\lower4pt\hbox{$\sim$}}}\hbox{$<$}}}$}}

\def\mnras{{MNRAS}}

\def\percm2{cm$^{-2}$}

\def\hii{\ion{H}{2}}

\received{}
\accepted{}
\slugcomment{To appear in {\it The Astrophysical Journal}.}
\shorttitle{AGNs in \hii\ Galaxies}
\shortauthors{Ulvestad \& Ho}
\journalid{}{}
\articleid{}{}

\begin{document}

\title{A Search for Active Galactic Nuclei in Sc Galaxies with \hii\ Spectra}

\author{James S.~Ulvestad}
\affil{National Radio Astronomy Observatory\altaffilmark{1},
P.O. Box O, Socorro, NM 87801} 
\email{julvesta@nrao.edu}
\and
\author{Luis C.~Ho}
\affil{The Observatories of the Carnegie Institution of Washington,
813 Santa Barbara St., Pasadena, CA 91011}
\email{lho@ociw.edu}

\altaffiltext{1}{The National Radio Astronomy Observatory is a facility of 
the National Science Foundation, operated under cooperative
agreement by Associated Universities, Inc.}

\begin{abstract}

We have searched for nuclear radio emission from a statistically complete
 sample of 40 
Sc galaxies within 30~Mpc that are optically classified as star-forming objects, in 
order to determine whether weak active galactic nuclei might be present.  Only 
three nuclear radio sources were detected, in NGC~864, NGC~4123, and 
NGC~4535.  These galaxies have peak 6-cm radio powers of
$\sim 10^{20}$~W~Hz$^{-1}$ at arcsecond resolution, while upper
limits of the non-detected galaxies typically range from 
$10^{18.4}$~W~Hz$^{-1}$ to $10^{20}$~W~Hz$^{-1}$.  The three nuclear 
radio sources all are resolved and appear to have diffuse
morphologies, with linear sizes of $\sim 300$~pc.  This
strongly indicates that circumnuclear star formation has been
detected in these three \hii\ galaxies.  
Comparison with previous 20-cm VLA results for the detected galaxies
shows that the extended nuclear radio emission has a flat spectrum
in two objects, and almost certainly is generated by thermal 
emission from gas ionized by young stars in the centers of those galaxies. 
The 6-cm radio powers are comparable to predictions for thermal 
emission that are based on the nuclear H$\alpha$ luminosities, and 
imply nuclear star formation rates of $0.08-0.8\,M_\odot$~yr$^{-1}$, 
while the low-resolution NRAO VLA Sky Survey implies galaxy-wide star 
formation rates of $0.3-1.0\,M_\odot$~yr$^{-1}$ in stars above 
$5M_\odot$.  In a few of the undetected
galaxies, the upper limits to the radio power are lower than 
predicted from the H$\alpha$ luminosity, possibly due to over-resolution
of central star-forming regions.  Although the presence 
of active nuclei powered by massive black holes cannot be definitively 
ruled out, the present results suggest that they are likely to be rare 
in these late-type galaxies with \hii\ spectra.

\end{abstract}


\section{Introduction}
\label{sec:intro}

Most nearby normal and active galaxies with bulges 
appear to have supermassive black holes at their centers (see
Kormendy \& Gebhardt 2001 for a review), and the masses
of these black holes ($\sim 10^6 M_\odot$--$10^9 M_\odot$)
seem well correlated with the galaxy
bulge masses as inferred from the bulge 
luminosities \citep{mag98}.  In fact, the relation of
black hole mass to bulge stellar velocity dispersion is a correlation with 
far less scatter than that derived from the bulge luminosity
\citep{fer00,geb00a,tre02}.  The black hole masses from
this relation are consistent with those derived from
the widths of broad emission lines and their distances from
the black hole as derived by reverberation mapping
\citep{geb00b,fer01}.  The relation of black hole mass to
bulge mass appears confirmed in at least one bulgeless
galaxy, the nearby spiral M33, where an upper limit of
1500--3000$M_\odot$ has been found for the black hole mass 
\citep{geb01,mer01b}.

Ho, Filippenko, \& Sargent (1995, 1997a, 1997b) used the Palomar 200 inch
telescope to perform a spectroscopic survey of 486 bright, nearby
galaxies selected from the Revised Shapley-Ames Catalog of 
Bright Galaxies \citep{san81}, in order to conduct a census
of the population of active galactic nuclei (AGNs) in the nearby universe.
Fifty-two Seyfert galaxies were found in that survey, and 82\% of 
the objects in a statistical sample of 45 of these Seyferts were detected
using the Very Large Array (VLA) at 6~cm, with observations 
of 15--18~min in length \citep{ho01,ulv01}.  The detected 
radio sources in these low-luminosity Seyferts typically are
associated with the AGNs, implying the presence of 
massive black holes powering the active nuclei.  

In addition to the 52 Seyferts identified by \citet{ho97a}, 
206 galaxies from the Palomar sample were reported to contain
\hii\ nuclei: galaxy nuclei with optical line ratios consistent
with ionization by young, massive stars.  As a general (though
not universal) rule, the AGNs (Seyferts and LINERs) are found in
early-type galaxies (earlier than Sbc), while the \hii\ nuclei
are found in galaxies of type Sb or later.  

Although the late-type galaxies in the Palomar sample that are
classified as \hii\ galaxies
appear to be dominated by star formation, it also is possible 
that they contain weak AGNs whose optical signature has simply been 
overpowered by the more dominant signal from nuclear \hii\ regions.  
The presence of
an AGN typically requires two constituents: a
massive central black hole and a supply of gas to ``feed''
that black hole.  Since
galaxies with \hii\ spectra generally harbor copious supplies of gas,
the question of whether \hii-dominated objects contain observable AGNs
then may be equivalent to the question of whether these
galaxies contain massive central black holes.  Good discriminators
for the presence of an AGN include the presence of
a compact source of hard X-rays or radio 

%\clearpage
%%%%%%%%%%%%%%%%%%%%%%%%%%%%%%%%%%%%%%%%%%%%%%%%%%%%%%%%%%%%%%%%%%%%%%%%%%
%good bounding box = 30 100 600 750
\begin{figure*}[t]
\centerline{\psfig{file=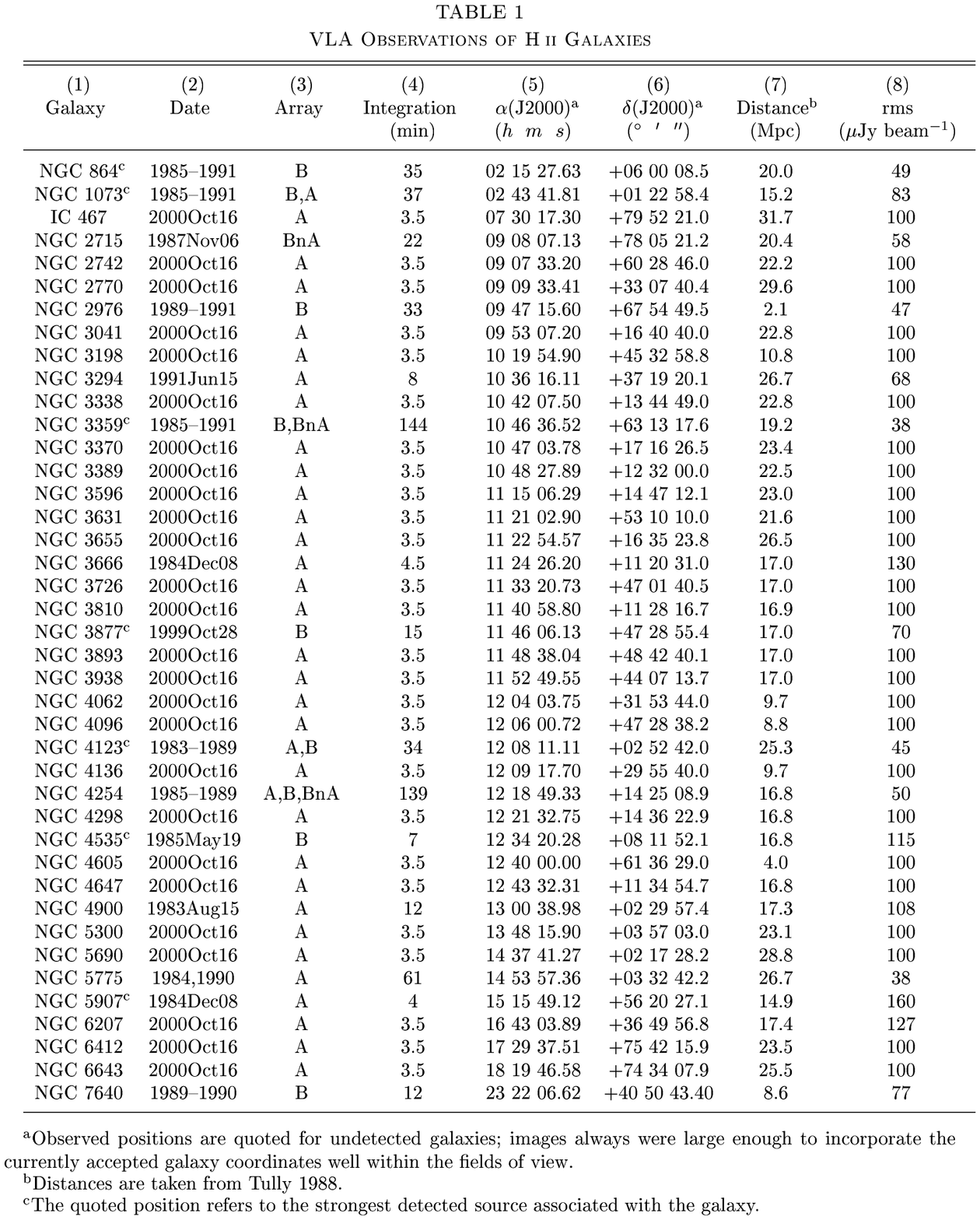,width=18.5cm,angle=0}}
\end{figure*}
%%%%%%%%%%%%%%%%%%%%%%%%%%%%%%%%%%%%%%%%%%%%%%%%%%%%%%%%%%%%%%%%%%%%%%%%%%

\noindent
emission identified
with the galaxy nucleus.  Here, we report on a study of a 
subsample of \hii\ nuclei in Sc galaxies from the
Palomar bright galaxy sample, in which we test for the presence
of weak AGNs and central black holes by searching for
radio emission from the galaxy nuclei.

\vspace{0.6cm}

\section{Sample Selection and Observations}
\label{sec:sample}

In order to provide a uniform sample,
we chose the 57 objects classified as Sc ($T$ = +5) in the
\citet{ho97a} catalog of \hii\ galaxies.  Due to the constraints
of limited observing time, we selected a distance-limited sample
of 41 of these galaxies lying within 30~Mpc.  In
this sample, 15 objects had 6-cm data

%%%%%%%%%%%%%%%%%%%%%%%%%%%%%%%%%%%%%%%%%%%%%%%%%%%%%%%%%%%%%%%%%%%%%%%%%%%
\vskip 0.3cm
%good bounding box = 180 5 430 750
\begin{figure*}[t]
\centerline{\psfig{file=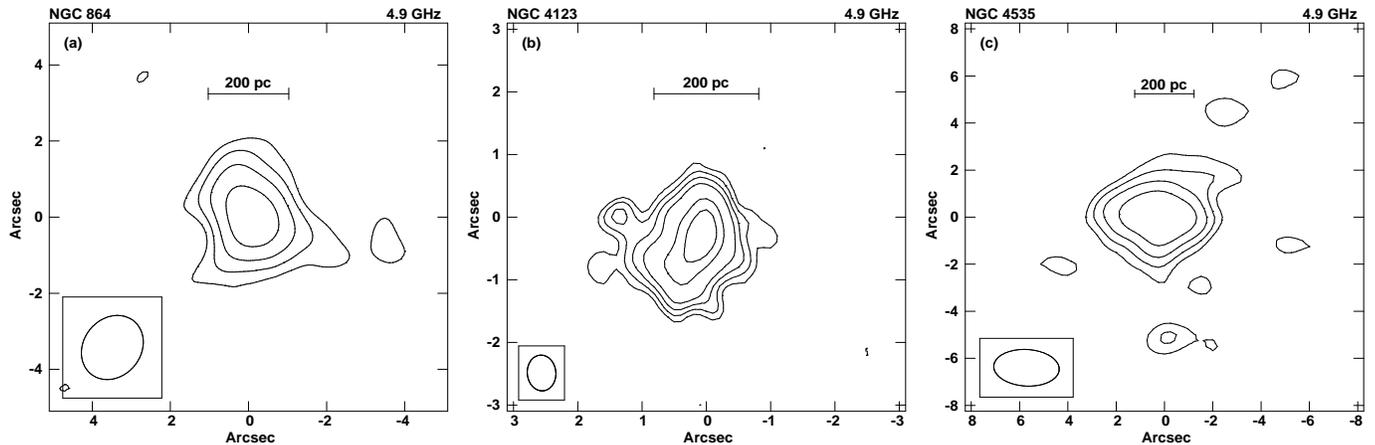,width=19.5cm,angle=270}}
\figcaption[f1.eps]{
Images at 4.9 GHz of the three detected nuclear radio
sources from the \hii\ galaxy sample.  All contour levels
start at 3 times the rms noise in the image, and increase
by factors of $\sqrt{2}$.  Scale bars indicating linear sizes of
200~pc are indicated on all images.  {\it (a)}  NGC 864: rms noise
of 49~$\mu$Jy~beam$^{-1}$, peak flux density of 0.55~mJy~beam$^{-1}$,
beam size 1\dsec 77$\times$1\dsec 48 at PA = $-35$\arcdeg.
{\it (b)} NGC 4123: rms noise of 45~$\mu$Jy~beam$^{-1}$, peak
flux density of 1.39~mJy~beam$^{-1}$,
beam size 0\dsec 57$\times$0\dsec 44 at PA = 5\arcdeg.
{\it (c)} NGC 4535: rms noise of 115~$\mu$Jy~beam$^{-1}$, peak
flux density of 1.66~mJy~beam$^{-1}$,
beam size 2\dsec 71$\times$1\dsec 55 at PA = 87\arcdeg.
\label{f1}}
\end{figure*}
\vskip 0.3cm
%%%%%%%%%%%%%%%%%%%%%%%%%%%%%%%%%%%%%%%%%%%%%%%%%%%%%%%%%%%%%%%%%%%%%%%%%%%

\noindent
at $\lesssim$1\arcsec\
resolution in the VLA archive, from observations
made in the high-resolution {\bf A} and {\bf B} configurations, or
the hybrid {\bf BnA} configuration with a long north
arm (see Thompson et al. 1980).  Data for these objects
were extracted from the archive.  Twenty-five sample galaxies,
as well as one galaxy at a distance slightly greater than 
30~Mpc (IC~467), were observed by us on 16 October 2000, 
using the VLA {\bf A} configuration.
All new observations were centered at a
frequency of 4.86~GHz, with a total bandwidth of 100~MHz in
each of two circular polarizations.  One galaxy that met our
selection criteria, NGC~3430, was not observed because of the
apparent presence of data in the VLA archive, but we later
found that there were no usable archive data.

Approximately 3.5 minutes were spent integrating on each of the 
newly observed galaxies, providing a detection threshold about
2.2 times higher than for the Seyfert galaxies observed by
\citet{ho01}.  For the archive data, total integration times ranged 
from a few minutes to more than two hours.  
Table~1 summarizes the new and archival observations of
the galaxies, including the nominal positions for the
galaxy nuclei that were used for imaging.  For the seven detected galaxies
(see Section~\ref{sec:results}), the listed positions are
those of the peaks of the radio sources apparently associated with the
galaxies.  All images covered fields of at least 100\arcsec\ in both
right ascension and declination, usually centered at the nominal galaxy 
positions at the times of the observations.  In two galaxies, NGC~3666
and NGC~5907, the currently accepted positions for the galaxy nuclei
are more than 15\arcsec\ from the original pointing positions, so
the data for these sources were phase-shifted to the more accurate 
galaxy positions in order to center the images on the nuclei.

Processing of both new and archival data sets was similar, and was
performed in the NRAO Astronomical Image Processing System 
\citep{van96}.  The
flux-density scales were set by observations of 3C~48 and 3C~286,
using the scale of \citet{baa77}, with slight
adjustments for variability that have been found by VLA staff.
A local calibrator was observed for each galaxy, in order to
transfer the amplitude scale from the primary flux calibrators
and to provide a local phase calibration, removing the
effects of electronics and atmosphere.  If more than one
data set was available from the archive for a given galaxy,
each data set was calibrated individually and then combined.
Bad data were identified and flagged, and
deconvolution and imaging then was performed on the calibrated
data sets.  All data sets were imaged with ``natural'' weighting
in the ({\it u,v\/}) plane in order to provide the best
possible sensitivity.  For data sets with excess noise
or obvious stripes in the images, confusing sources were imaged
and self-calibrated in order to reduce the noise level.
The $1\sigma$ rms~noise was 100~$\mu$Jy~beam$^{-1}$ for
all the new observations, and varied depending on data quality
and quantity, as well as array configuration for the archival data 
(see Table~1).  Synthesized
beam sizes were 0\dsec 4--0\dsec 5 for the new {\bf A} 
configuration data, and 0\dsec 4 to as much as 2\arcsec\ for the 
archival data.

\section{Results}
\label{sec:results}

The range of resolutions and galaxy distances (typically 15--30~Mpc)
provides linear resolutions in the range of 40 to 300~pc, 
sufficient to isolate the galaxy center from the
rest of the galaxy.  None of the 26 newly observed galaxies was
detected, with a $5\sigma$ upper limit of 0.5~mJy~beam$^{-1}$
for each.  For the 15 galaxies whose archival data were analyzed, seven were 
detected: NGC~864, 1073, 3359, 3877, 4123, 4535, and 5907.  The flux densities 
and powers of the detected galaxies are summarized in 
Table~2, while the positions of their radio
peaks were given in Table~1.

The noise levels for the archival data often were lower than 
for the new data due to longer integration times, but only one
of the detections was weaker than the upper limit for the
new data.  However, the archival galaxies often contain significant amounts
of data from the larger {\bf B} configuration, sensitive to more extended 
structures of lower surface brightness than for the new {\bf A}
configuration data.  The beam area in the larger configuration
typically is 10 times larger than the beam area in the smaller
configuration; the respective beam diameters are $\sim 130$~pc
and $\sim 40$~pc at the median galaxy distance of 17~Mpc.  If the 
radio sources tend to be resolved on scales of 100--200~pc, an increased 
peak flux density would be seen in the lower resolution
data, perhaps moving the galaxies above our detection 
threshold.  This possibility is supported by the fact that 
the detected nuclear radio sources in three galaxies
are partly resolved 

%%%%%%%%%%%%%%%%%%%%%%%%%%%%%%%%%%%%%%%%%%%%%%%%%%%%%%%%%%%%%%%%%%%%%%%%%%
%good bounding box = 10 320 600 550
\begin{figure*}[t]
\centerline{\psfig{file=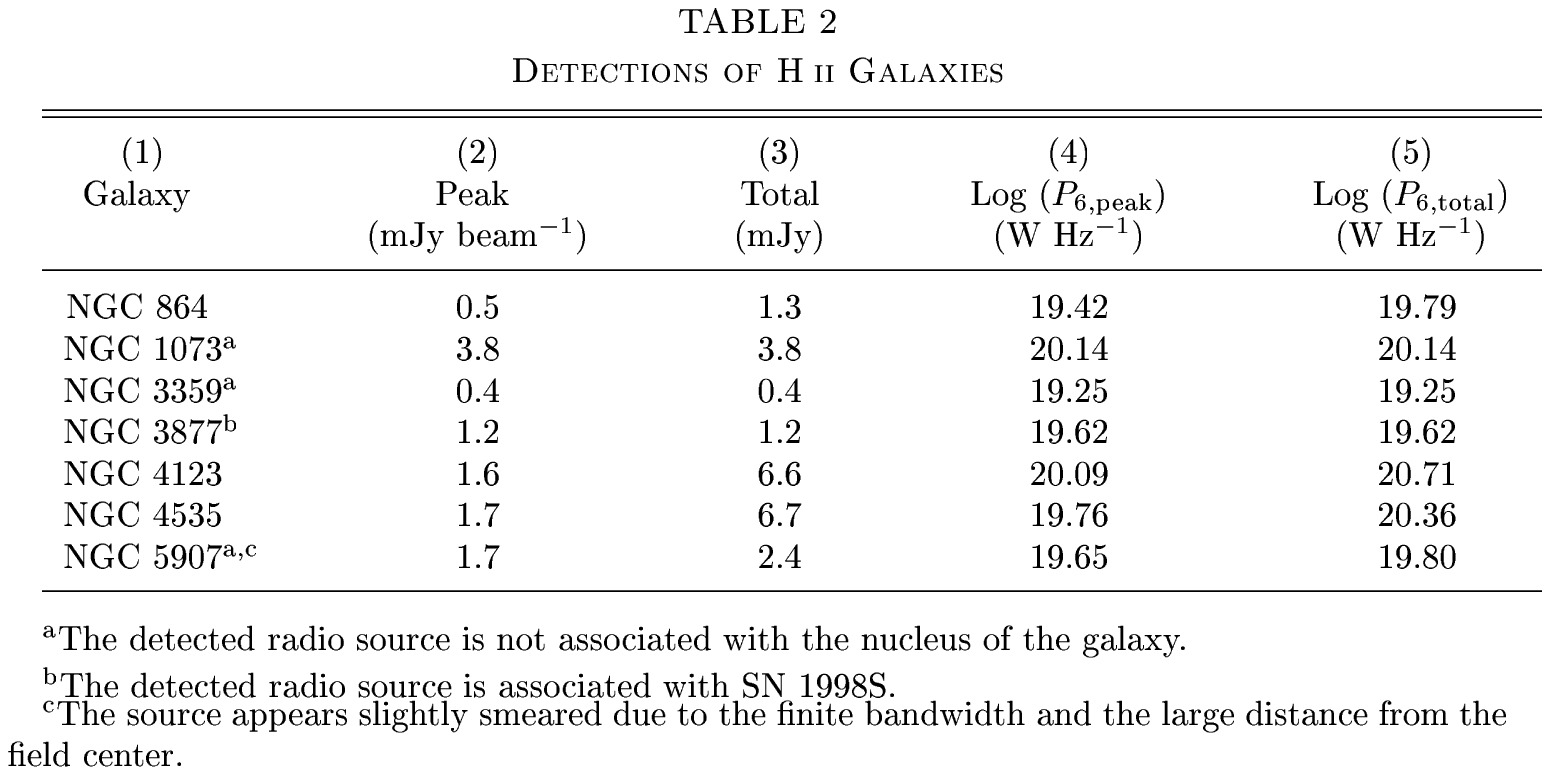,width=18.5cm,angle=0}}
\end{figure*}
%%%%%%%%%%%%%%%%%%%%%%%%%%%%%%%%%%%%%%%%%%%%%%%%%%%%%%%%%%%%%%%%%%%%%%%%%%

\noindent
by the VLA (see Section~\ref{sec:locations} for
more details).

\vspace{0.3cm}

\section{Origins of the Radio Emission}
\label{sec:origins}

\subsection{Locations of the Detected Radio Sources}
\label{sec:locations}

In this section, we consider primarily the \hii\ galaxies
that were detected by the VLA.  Only three of
those galaxies have radio detections consistent with the
positions of their nuclei as given in the NASA Extragalactic
Database (NED) and confirmed with the measurements of
\citet{cot99}.  The three nuclear detections are in the
galaxies 

%%%%%%%%%%%%%%%%%%%%%%%%%%%%%%%%%%%%%%%%%%%%%%%%%%%%%%%%%%%%%%%%%%%%%%%%%%
\vskip 0.3cm
%good bounding box = 36     94    576    697

\psfig{file=f2.eps,width=8.5cm,angle=0}
\figcaption[f2.eps]{
VLA image of NGC~253, from \citet{ulv97}, degraded
in resolution and flux density to indicate its appearance viewed
at a distance of 20~Mpc.  Contours increase by $\sqrt{2}$ from
a lowest level of 150~$\mu$Jy~beam$^{-1}$.  At 20 Mpc distance,
the beam size is 0\dsec 66$\times$0\dsec 36 at PA = 12\arcdeg,
and the peak flux density is 8.22~mJy~beam$^{-1}$.
\label{f2}}
\vskip 0.3cm
%%%%%%%%%%%%%%%%%%%%%%%%%%%%%%%%%%%%%%%%%%%%%%%%%%%%%%%%%%%%%%%%%%%%%%%%%%

\noindent
NGC~864, NGC~4123, and NGC~4535.  NGC~1073 has
two sources detected in the field, one associated with a
background quasar identified by \citet{col02}; the other
detection may be associated with the galaxy, but is located
approximately 30\arcsec\ from the nucleus at position angle
37\deg, and is not coincident with either of the
intermediate-luminosity X-ray objects shown by \citet{col02}.
NGC~3359 has a detection about 9\dsec 5 south of the
nucleus, and NGC~5907 has a detection nearly an arcminute away
from the nucleus, possibly associated with the disk of the
edge-on host galaxy.  Finally, the only source detected in NGC~3877 is
at the location of the Type IIn supernova SN~1998S, whose
radio detection using the same data was reported previously
by \citet{van99}.

Figure~1 shows the radio images of the three
nuclear radio sources found in our survey.  All three are
resolved and appear somewhat diffuse (as opposed to 
``jet-like''), with diameters of $\sim 300$~parsecs.  This radio 
emission may be associated with starbursts, but the unresolved 
components of the core radio emission conceivably may be 
associated with AGNs.  For these three nuclear radio
sources, NGC~864 and NGC~4535 contain only data from the
{\bf B} configuration of the VLA, whereas NGC~4123 has data
from both the {\bf A} and {\bf B} configurations.  
To test the effects of resolution on galaxy
detectability, we have made images of NGC~4123 containing
only the {\bf A} configuration data and only the {\bf B}
configuration data.  The peak flux density of the nuclear
source is a factor of 3 higher in the {\bf B} configuration
image than in the {\bf A} configuration image (4.4~mJy~beam$^{-1}$
{\it vs.} 1.4~mJy~beam$^{-1}$), as expected from inspection
of Figure~1.  If the same ratio had held for NGC~864
and NGC~4535, NGC~864 would not have been detected in our short
{\bf A} configuration integrations, whereas NGC~4535 would have
been marginally detected.  Therefore, the generally higher 
sensitivity to extended structures in the archival galaxies accounts
for at least some of the increased detection probability for
these objects.  In fact, it may be that a significant fraction
of the galaxies that were not detected in the October 2000 observing
run would have been detected using a more compact VLA configuration.

\subsection{Radio Emission Associated with Star Formation}
\label{sec:sf}

A young starburst typically can give rise to thermal radio 
emission from collections of \hii\ regions or super
star clusters, 

%%%%%%%%%%%%%%%%%%%%%%%%%%%%%%%%%%%%%%%%%%%%%%%%%%%%%%%%%%%%%%%%%%%%%%%%%%
%good bounding box = 10 340 600 530
\begin{figure*}[t]
\centerline{\psfig{file=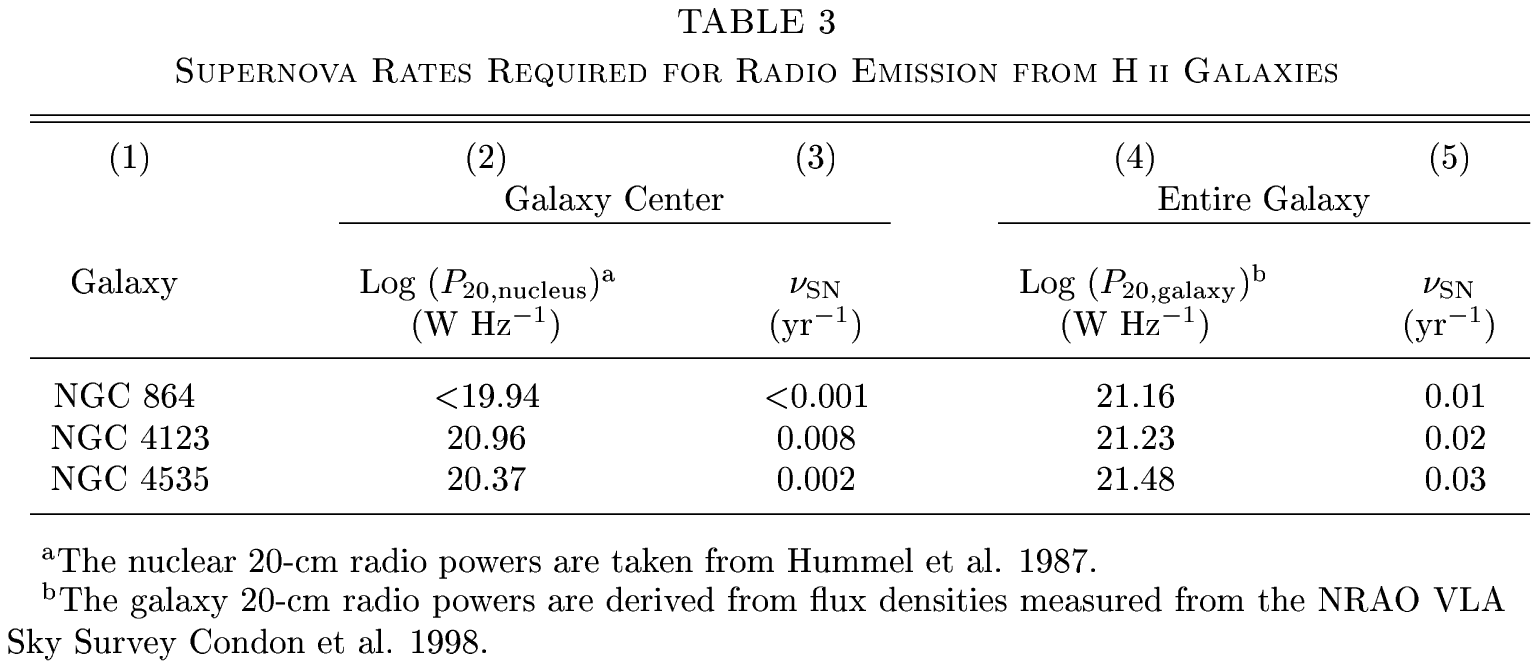,width=18.5cm,angle=0}}
\end{figure*}
%%%%%%%%%%%%%%%%%%%%%%%%%%%%%%%%%%%%%%%%%%%%%%%%%%%%%%%%%%%%%%%%%%%%%%%%%%

\noindent
or to nonthermal radio emission from
collections of supernova remnants.  Such radio emission,
with varying ratios of thermal and nonthermal components,
is observed in starburst galaxies such as NGC~253 (Antonucci \& Ulvestad 1988;
Ulvestad \& Antonucci 1997; Johnson et al. 2001), M82 \citep{kro85,mux94}, 
Henize 2-10 \citep{kob99,bec01}, and NGC~5253 \citep{tur00,gor01}, 
or in merger galaxies such as 
NGC~4038/9 \citep{nef00}.  The three \hii\ galaxies with nuclear
radio detections that are shown in Figure~1
have radio sizes of a few hundred parsecs, indicating that
any intense star formation must be confined to the near-nuclear regions
of the galaxies.  

%%%%%%%%%%%%%%%%%%%%%%%%%%%%%%%%%%%%%%%%%%%%%%%%%%%%%%%%%%%%%%%%%%%%%%%%%%
\vskip 0.3cm
%good bounding box = 36     94    576    697

\psfig{file=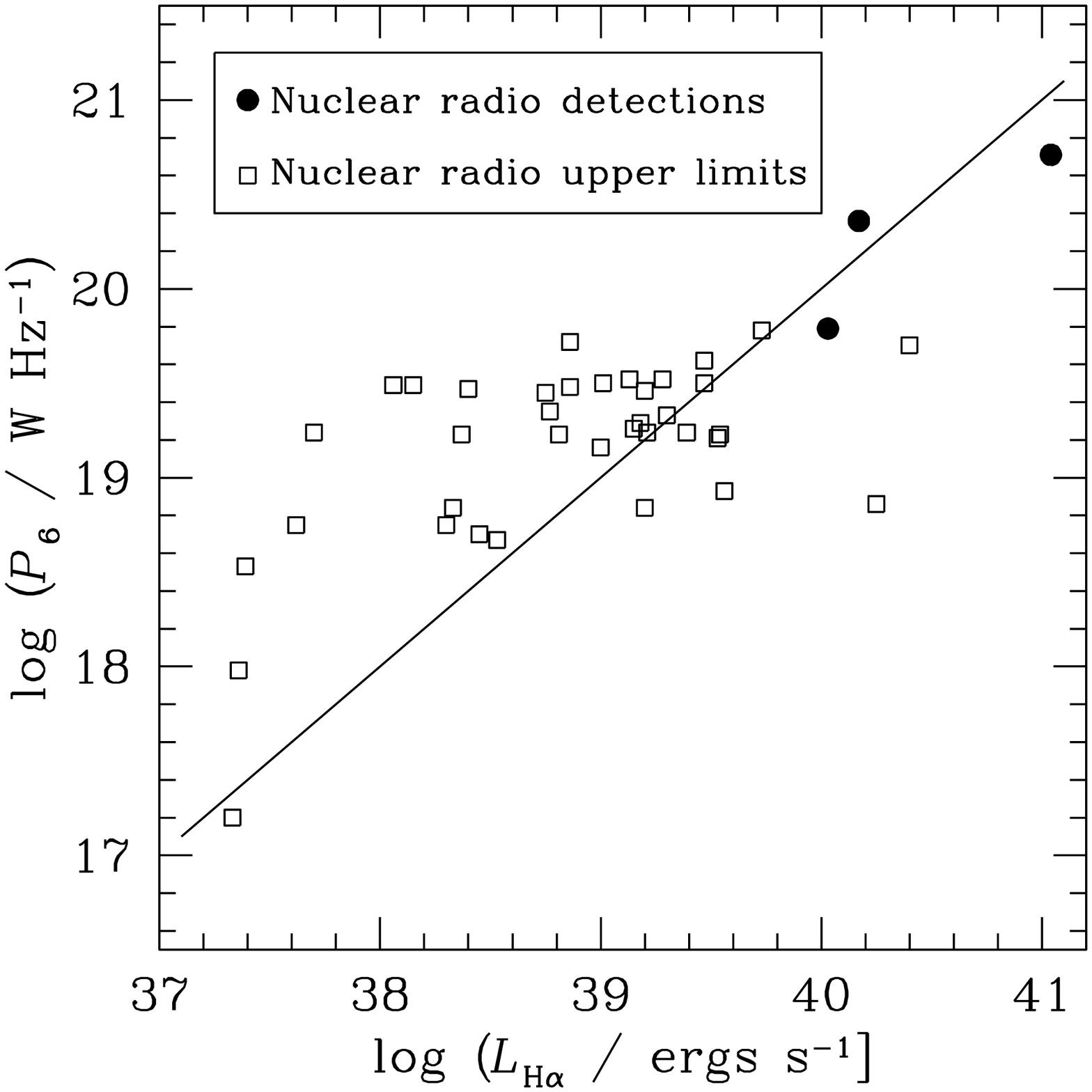,width=8.5cm,angle=0}
\figcaption[f3.eps]{
Plot of radio power vs. H$\alpha$ luminosity for 40
Sc galaxies with \hii\ nuclei in the statistical sample.  Radio
detections and upper limits are distinguished by different symbols.
The diagonal line is a prediction of thermal radio emission based on
the H$\alpha$ luminosity, using $P_6\ ({\rm W\ Hz}^{-1})\approx
10^{-20}L_{{\rm H}\alpha}\ ({\rm ergs\ s}^{-1})$ (Filho et al. 2002).
\label{f3}}
\vskip 0.3cm
%%%%%%%%%%%%%%%%%%%%%%%%%%%%%%%%%%%%%%%%%%%%%%%%%%%%%%%%%%%%%%%%%%%%%%%%%%%

The 4.9-GHz radio sizes of NGC~864, NGC~4123, and NGC~4535
are quite similar to the 500--600~pc nuclear starburst in the nearby Sc
galaxy NGC~253 \citep{ant88,ulv97}.  NGC~253 has a total 5-GHz flux
density of 2.4~Jy integrated over the entire galaxy \citep{gri94}.
However, the flux density found by integrating over the central 
NGC~253 starburst in the 4.9-GHz image shown by \citet{ulv97} is only
1.3~Jy, corresponding to a radio power
of $9.6\times 10^{20}$~W~Hz$^{-1}$.  The three \hii\ nuclei whose
detections are reported here have radio powers ranging from
a factor of 2 (for NGC~4123) to a factor of 15 (for NGC~864)
lower than this central starburst in NGC~253.  Thus, it is entirely 
possible that the detected \hii\ galaxies harbor nuclear star formation
somewhat weaker than that in NGC~253.  Supporting this possibility, 
Figure~2 shows the 6-cm image of NGC~253 \citep{ulv97}, 
as it would appear at a distance of 20~Mpc instead of at 2.5~Mpc, having 
the flux density reduced by a factor of 64, and
angular resolution degraded by a factor of 8 in each dimension.
The radio morphology of this degraded NGC~253 image looks 
similar to the \hii\ galaxies in Figure~1, except that the 
respective peak and total flux densities are somewhat higher, 
at 8.2~mJy~beam$^{-1}$ and 20~mJy.

The archival 6-cm data for NGC~864 and NGC~4535 were taken in the
{\bf B} configuration, and we also have measured a flux density from the
{\bf B} configuration 6-cm data for NGC~4123 (total flux density given in
Table~2).  Comparing the 6-cm values with the 20-cm {\bf A} 
configuration flux densities (at similar angular resolution) published 
by \citet{hum87} indicates that the resolved nuclear sources in NGC~4123
and NGC~4535 have respective two-point spectral indices 
(using $S_\nu\propto\nu^{+\alpha}$) of $-0.51$ and $-0.02$; due to its
non-detection at 20~cm, the spectral index of NGC~864 is limited to a
value greater than $-0.27$.
The relatively flat spectra of at least two of the objects may be indicative
of radio emission dominated by optically thin thermal emission from
star formation regions.

For a thermal radio source, both the hydrogen emission lines and the free-free
radio emission are powered by ionizing photons from 
hot young stars.  The radio power and H$\alpha$
luminosity are related by \citep{fil02}
\begin{equation}
P_6\ ({\rm W\ Hz}^{-1})\ \approx 10^{-20} L_{{\rm H}\alpha}
\ ({\rm ergs\ s}^{-1})\ .
\label{eqn:halpha}
\end{equation}
For all 40 \hii\ galaxies observed in our distance-limited 
sample (those in Table~1 except for IC~467), 
Figure~3 plots the radio power against the 
H$\alpha$ luminosity given by \citet{ho97a}\footnote{As
explained in Ho, Filippenko, \& Sargent (2002), some of the H$\alpha$
luminosities published in Ho et al. (1997a) have been updated with more 
accurate values from the literature.  These are listed in Ho et al. (2002).}.
This figure shows that three of the five galaxies with the highest

%%%%%%%%%%%%%%%%%%%%%%%%%%%%%%%%%%%%%%%%%%%%%%%%%%%%%%%%%%%%%%%%%%%%%%%%%%
\vskip 0.3cm
%good bounding box = 80 150 490 680

\psfig{file=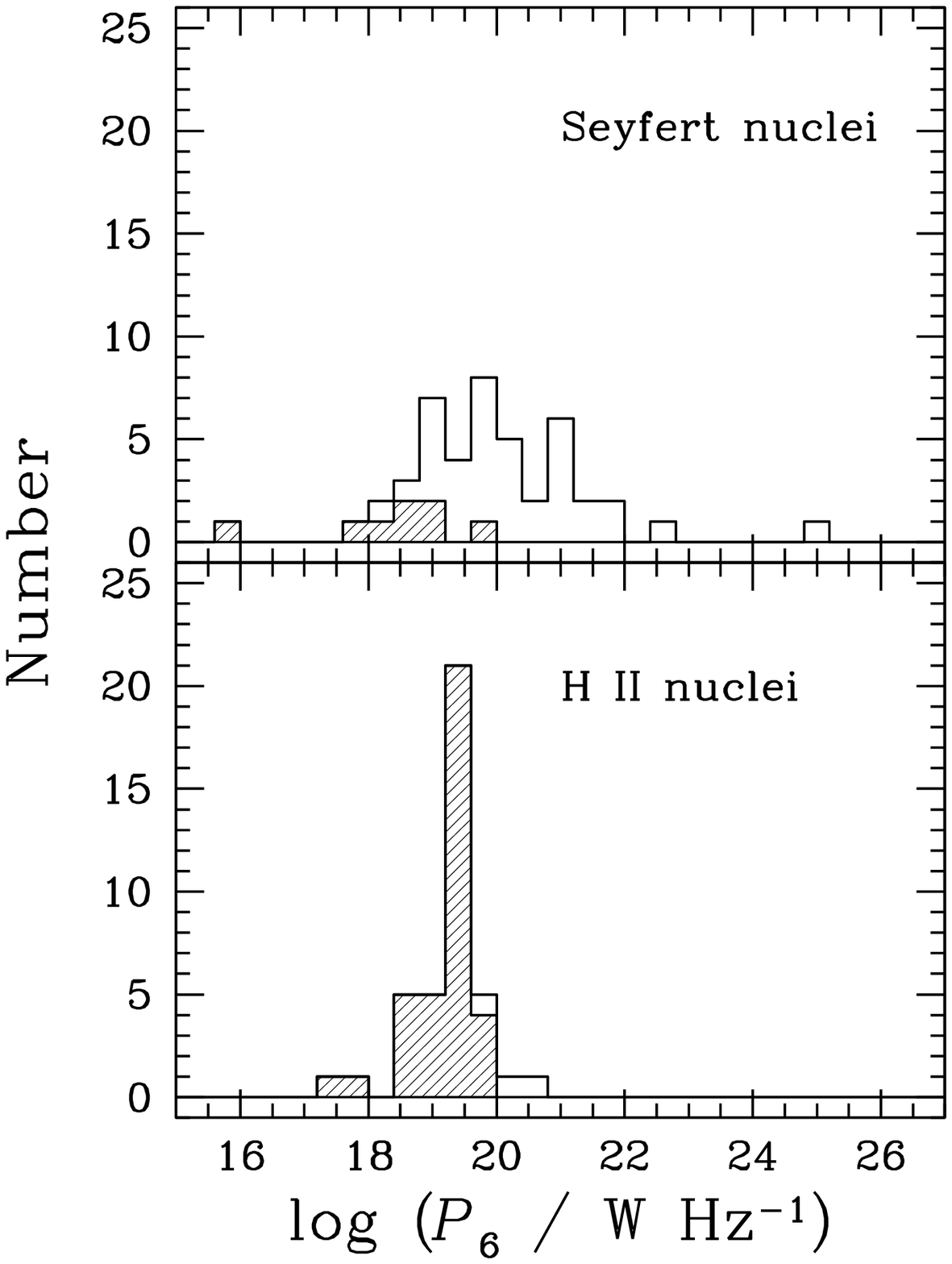,width=8.5cm,angle=0}
\figcaption[f4.eps]{
Histogram of radio powers of 40 Sc galaxies
with \hii\ nuclei within 30~Mpc,
compared to Seyfert galaxies from the same Palomar
galaxy sample.  Powers for the Seyfert galaxies have
been taken from \citet{ulv01}.  Upper limits are shown
hatched.
\label{f4}}
\vskip 0.3cm
%%%%%%%%%%%%%%%%%%%%%%%%%%%%%%%%%%%%%%%%%%%%%%%%%%%%%%%%%%%%%%%%%%%%%%%%%%%

\noindent
H$\alpha$ luminosities harbor detected nuclear radio sources.
Furthermore, these three detected galaxies fall near
the relation given by Equation~\ref{eqn:halpha}, supporting the
inference (from their spectra) that their radio emission has a significant
thermal component.  We note that the radio upper limits for some of the
undetected galaxies actually are below
the predictions based on the nuclear H$\alpha$ luminosity,
with the largest discrepancy existing for NGC~3877.  Since
the H$\alpha$ luminosities were measured with a 2\arcsec\ by
4\arcsec\ aperture, it may be that nuclear thermal radio sources have
been over-resolved by the higher resolution VLA imaging.

If the radio emission is thermal and related to starbursts,
the inferred star formation rate also can be computed.
\citet{ken98} gives the following formula for the star
formation rate inferred from the H$\alpha$ luminosity, 
assuming a Salpeter mass function for 0.1--100$M_\odot$
and solar metallicity:
\begin{equation}
{\rm SFR}\ (M_\odot\ {\rm yr}^{-1})\ =\ 7.9 \times 10^{-42}\ 
L({\rm H}\alpha)\ ({\rm ergs\ s}^{-1})\ .
\label{eqn:halpha-sfr}
\end{equation}
For our three detected nuclei, the H$\alpha$ luminosities
range from $10^{40}$~ergs~s$^{-1}$ to $10^{41}$~ergs~s$^{-1}$
so the inferred nuclear star formation rates that could account for 
the H$\alpha$ luminosity would be 
$0.08-0.8\,M_\odot$~yr$^{-1}$.  Alternatively, using the formalism
given by \citet{con92}, the inferred star formation rates in stars 
above $5M_\odot$ are only $0.02-0.2\,M_\odot$~yr$^{-1}$.

Another possible result of a circumnuclear starburst would
be a compact radio source with a steep spectrum, 
made up of complexes of supernova remnants that are
unresolved by the VLA.  Examples of these have been
imaged at higher resolution by the Very Long Baseline
Array (VLBA), in Arp~220 \citep{smi98} and Mrk~231 \citep{tay99}.
The total radio powers can be used to estimate the associated
supernova rates that would be required to account for the radio 
sources.  
%If we assume that the 
%nuclear radio emission is directly associated with supernova 
%remnants and that the emitting electrons have a spectral 
%index of $-0.5$, the relation between the 20-cm radio power and the
%supernova rate $\nu_{\rm SN}$ is \citep{ulv82,con92}
%\begin{equation}
%\biggl({P_{20}\over 10^{22}\ {\rm W\ Hz}^{-1}}\biggr)
%\sim 1.0\ \biggl({\nu_{\rm SN}\over {\rm yr}^{-1}}\biggr)\ .
%\label{eqn:jsusnr}
%\end{equation}
%After conversion of the total 6-cm radio powers (Table~2)
%to 20-cm powers using a spectral index of $-0.5$, 
%the nuclear supernova rates necessary to account for the observed
%radio emission range from 0.01~yr$^{-1}$ 
%(NGC~864) to 0.09~yr$^{-1}$ (NGC~4123).
We use the formalism for an object whose total radio emission
is dominated by cosmic rays accelerated in supernova remnants,
but not identifiable with specific remnants.  In this case,
for a spectral index of $-0.5$, Equation~(18) of 
\citet{con92} becomes
\begin{equation}
\Biggl({P_{20}\over 10^{22}\ {\rm W\ Hz}^{-1}}\Biggr)
\sim 11\ \Biggl({\nu_{\rm SN}\over {\rm yr}^{-1}}\Biggr)\ .
\label{eqn:jjcsnr}
\end{equation}
For the nuclear radio sources, we can use the 20-cm radio emission
found by \citet{hum87} to estimate the nuclear supernova rates if the
nuclear emission were entirely nonthermal (although the spectra
appear thermal in two cases), and find values less than 0.01~yr$^{-1}$
for all three galaxies.  We also can compute the total supernova rates 
for the entire galaxies by using the galaxy flux densities determined from the 
NRAO VLA Sky Survey \citep{con98}; this implies required supernova
rates for the galaxies ranging from 0.01~yr$^{-1}$ (NGC~864)
to 0.03~yr$^{-1}$ (NGC~4535).  Results of the two different
computations are summarized in Table~3.
The star formation rates in stars above $5M_\odot$ \citep{con92}
required to account for the radio emission (presumed nonthermal) for
the entire galaxies range from $\sim 0.3M_\odot$~yr$^{-1}$ to 
$\sim 1M_\odot$~yr$^{-1}$.  These values are roughly an 
order of magnitude higher than those required for the nuclear sources,
whether they are dominated by thermal or nonthermal emission.
However, they are consistent with predictions for the entire galaxies 
that are based on the far-infrared fluxes measured by the {\it Infrared
Astronomical Satellite} \citep{soi89,mos90}, converted to star 
formation rates using the formulation summarized by \citet{ken98}.

\subsection{Radio Emission Associated with Active Galactic Nuclei}
\label{sec:agns}

The other possibility for the radio sources in the \hii\ nuclei 
of Sc galaxies is that they have radio cores associated with
AGNs and supermassive black holes.  There
has been considerable discussion in the literature about whether
the core radio emission might be correlated with black hole
masses in these cases (e.g., Franceschini, Vercellone, \& Fabian 1998; 
Laor 2000).  In fact, \citet{ho02}
contends that any correlation for weakly active galaxies
is an indirect consequence
of a more fundamental relationship between radio luminosity and
optical bulge luminosity (or mass) as well as the established
relationship between bulge luminosity and black hole mass.

Since Sc galaxies such as those in our \hii\ sample 
generally have no well-defined bulges whose luminosities
can be measured, and no known black holes, direct predictions
of radio power cannot be made based on such properties.  Instead,
as a comparison, we have plotted a histogram of 6-cm nuclear
radio powers for the 40 \hii\ galaxies in our statistical sample
in Figure~4, and compared the results to a
similar plot for the statistical sample of 45 Seyfert nuclei
(39 with galaxy types earlier than Sc)
that was published by \citet{ulv01}.  Application of the
Gehan's Generalized Wilcoxon Test \citep{iso90,lav92} indicates
that the probability that the Seyferts and \hii\ galaxies were
drawn from the same parent population in radio luminosity is
less than $10^{-4}$.  This result would be expected even if
the \hii\ galaxies contain AGNs, since they have faint or non-existent
bulges, and there are no active galaxies with faint bulges
that have high radio powers \citep{ho02}.  

In general, the radio sources in low-luminosity AGNs are quite
compact.  For the Seyferts from the Palomar sample, \citet{ho01}
showed that about half the objects have diameters of $\sim 100$~pc
or less.  In the case of LINERs, \citet{nag00} demonstrated that
those with flat radio spectra typically have AGN radio cores with diameters
less than 25~pc, and \citet{fal00} used the VLBA to show that these cores
typically are parsec-scale or smaller.  The radio peaks
of NGC~864, NGC~4123, and NGC~4535,
having size limits less than 50--150~pc, are consistent with the presence 
of AGNs but are not required to be black-hole powered.  Instead, the trend of
decreasing peak flux density with higher resolution seen in NGC~4123
(see Section~\ref{sec:locations})
indicates that those peaks probably would be resolved if somewhat smaller
spatial scales (longer baselines) were sampled.  If so, AGNs are unlikely 
to contribute much to the currently unresolved radio emission.
%\subsection{Comparison of AGN and Starburst Origins for Radio Emission}
%\label{sec:comp}
%
%The presence of AGNs in the Sc galaxies having \hii\ spectra cannot 
%be ruled out based solely on the radio emission.  However, 
%the large number of upper limits and the morphology of the
%detected sources gives no indication that massive black holes
%associated with nuclei that predominantly show \hii\ spectra
%can power a substantial amount of radio emission in those galaxies.  
%Instead, nuclear star formation seems a far more likely explanation 
%for the three detected galaxy nuclei.  
If there really are weak AGNs 
causing the most compact radio emission, they should be detectable with 
the VLBA, as are many other low-luminosity AGNs \citep{fal00,nag02}.

\section{Summary}

We have searched for nuclear radio sources in a statistical sample of 40 Sc 
galaxies hosting \hii\ nuclei, using imaging of new and archival VLA
data at 6~cm.  Only three such sources were 
detected, in galaxies among those with the largest H$\alpha$
luminosities in the sample.  The radio  powers and morphologies
are consistent with the association of the detected radio sources
with nuclear starbursts similar to that in NGC~253, rather than
being caused by active galaxies powered by massive black holes.
The ratio of radio power to H$\alpha$ luminosity 
is generally consistent with thermal radio emission from nuclear starbursts, and
the detected central sources could produce such thermal emission for star
formation rates of only $0.08-0.8\,M_\odot$~yr$^{-1}$.
However, nonthermal emission from supernova remnants also could
contribute to the detected nuclear sources, requiring supernova rates
of well under  0.01~yr$^{-1}$ and star formation rates of 
$\sim 0.1M_\odot$~yr$^{-1}$.  

To the extent that radio cores are a common feature in low-luminosity AGNs, 
the nondetection of radio cores in \hii\ nuclei strongly suggests that they 
intrinsically lack AGNs, and, by inference, massive black holes.  This is not 
to say that late-type galaxies never host AGNs.  Indeed, $\sim$15\% of the 
Seyfert nuclei and $\sim$10\% of all AGNs in the Palomar survey are hosted in 
galaxies with Hubble types Sc or later.  In terms of AGN demographics, the 
present results imply that the existing statistics of AGNs in late-type 
galaxies based on optical searches (Ho et al. 1997b) are likely to be robust. 
There is not a large population of AGNs hidden among late-type galaxies
that have \hii\ spectra.

\acknowledgments

We thank the anonymous referee for extremely useful comments that
helped clarify and focus the paper.
This research has made use of the NASA/IPAC Extragalactic Database (NED) which 
is operated by the Jet Propulsion Laboratory, California Institute of 
Technology, under contract with the National Aeronautics 
and Space Administration.  In addition, this research has made use of NASA's
Astrophysics Data System Abstract Service.  The work of L.~C.~H. is partly 
funded by NASA grants awarded by the Space Telescope Science Institute, which 
is operated by AURA, Inc., under NASA contract NAS5-26555. 

%\clearpage


\begin{thebibliography}{}

\bibitem[Antonucci \& Ulvestad(1988)]{ant88} Antonucci, R. R. J.,
\& Ulvestad, J. S. 1988, \apjl, 330, L97

\bibitem[Baars et al.(1977)]{baa77} Baars, J. W. M., Genzel, R.,
Pauliny-Toth, I. I. K., \& Witzel, A. 1977, \aap, 61, 99

\bibitem[Beck, Turner \& Gorjian(2001)]{bec01} Beck, S. C.,
Turner, J. L., \& Gorjian, V. 2001, \aj, 122, 1365

\bibitem[Colbert \& Ptak(2002)]{col02} Colbert, E. J. M., \&
Ptak, A. F. 2002, \apjs, in press (astro-ph/0204002)

\bibitem[Condon(1992)]{con92} Condon, J. J. 1992, \araa, 20, 575

\bibitem[Condon et al.(1998)]{con98} Condon, J. J., Cotton, W. D., Greisen, E. W.,
Yin, Q. F., Perley, R. A., Taylor, G. B., \& Broderick, J. J.
1998, \aj, 115, 1693

\bibitem[Cotton, Condon, \& Arbizzani(1999)]{cot99} Cotton, W. D.,
Condon, J. J., \& Arbizzani, E. 1999, \apjs, 125, 409

\bibitem[Falcke et al.(2000)]{fal00} Falcke, H.,
Nagar, N. M., Wilson, A. S., \& Ulvestad, J. S.
2000, \apj, 542, 197

\bibitem[Ferrarese \& Merritt(2000)]{fer00} Ferrarese, L. \& Merritt, D. 
2000, \apjl, 539, L9

\bibitem[Ferrarese et al.(2001)]{fer01} Ferrarese, L., Pogge, R. W.,
Peterson, B. M., Merritt, D., Wandel, A., \& Joseph, C. L. 2001,
\apjl, 555, L79

\bibitem[Filho, Barthel, \& Ho(2002)]{fil02} Filho, M. E., Barthel, P. D.,
\& Ho, L. C. 2002, \apjs, in press (astro-ph/0205196)

\bibitem[Franceschini, Vercellone, \& Fabian(1998)]{fra98}
Franceschini, A., Vercellone, S., \& Fabian, A. C. 1998, \mnras, 297, 817

\bibitem[Gebhardt et al.(2000a)]{geb00a} Gebhardt, K., et al. 2000a, \apjl, 539, L13

\bibitem[Gebhardt et al.(2000b)]{geb00b} ------. 2000b, \apjl, 543, L5

\bibitem[Gebhardt et al.(2001)]{geb01} ------. 2001,
\aj, 122, 2469

\bibitem[Gorjian, Turner, \& Beck(2001)]{gor01} Gorjian, V., Turner, J. L.,
\& Beck, S. C. 2001, \apjl, 554, L29

\bibitem[Griffith et al.(1994)]{gri94} Griffith, M. R., Wright, A. E.,
Burke, B. F., \& Ekers, R. D. 1994, \apjs, 90, 179

%\bibitem[Ho(1999)]{ho99} Ho, L. C. 1999, \apj, 516, 672

\bibitem[Ho(2002)]{ho02} Ho, L. C. 2002, \apj, 564, 120

\bibitem[Ho et al.(1995)]{ho95} Ho, L. C., 
Filippenko, A. V., \& Sargent, W. L. W. 1995, \apjs, 98, 477

\bibitem[Ho et al.(1997a)]{ho97a} ------.  1997a, \apjs, 112, 315

\bibitem[Ho et al.(1997b)]{ho97b} ------.  1997b, \apj, 487, 568

\bibitem[Ho et al.(2002)]{ho02b} ------.  2002, \apj, submitted

\bibitem[Ho \& Ulvestad(2001)]{ho01} Ho, L. C., \& Ulvestad, J. S. 2001, 
\apjs, 133, 77

\bibitem[Hummel et al.(1987)]{hum87} Hummel, E., van der Hulst, J. M.,
Keel, W. C., \& Kennicutt, R. C., Jr. 1987, \aaps, 70, 517

\bibitem[Isobe \& Feigelson(1990)]{iso90} Isobe, T., \&
Feigelson, E. D. 1990, \baas, 22, 917

\bibitem[Johnson et al.(2001)]{joh01} Johnson, K. E., Kobulnicky, H. A.,
Massey, P., \& Conti, P. S. 2001, \apj, 559, 864

\bibitem[Kennicutt(1998)]{ken98} Kennicutt, R. C., Jr. 1998,
\araa, 36, 189

\bibitem[Kobulnicky \& Johnson(1999)]{kob99} Kobulnicky, H. A., \&
Johnson, K. E. 1999, \apj, 527, 154

\bibitem[Kormendy \& Gebhardt(2001)]{kor01} Kormendy, J., \& Gebhardt, K.
2001, in 20th Texas Symposium on Relativistic Astrophysics, AIP Conf.
Proceedings 586, ed. J. C. Wheeler \& H. Martel (New York: American
Institute of Physics), 363

\bibitem[Kronberg, Biermann, \& Schwab(1985)]{kro85}
Kronberg, P. P., Biermann, P., \& Schwab, F. R. 1985, \apj, 291, 693

\bibitem[Laor(2000)]{lao00} Laor, A. 2000, \apjl, 543, L111

\bibitem[LaValley, Isobe, \& Feigelson(1992)]{lav92} LaValley, M.,
Isobe, T., \& Feigelson, E. D. 1992, in ASP Conf. Ser. 25,
Astronomical Data Analysis Software and Systems I, ed. D. M. Worrall,
C. Biemesderfer \& J. Barnes (San Francisco: ASP), 245

\bibitem[Magorrian et al.(1998)]{mag98} Magorrian, J., et al. 1998,
\aj, 115, 2285

%\bibitem[Merritt \& Ferrarese(2001)]{mer01} Merritt, D., \&
%Ferrarese, L. 2001, in The Central Kiloparsec of Starbursts and AGN:
%The La Palma Connection,
%ASP Conf. Ser. 249, ed. J. H. Knapen, J. E. Beckman, I. Shlosman,
%\& T. J. Mahoney (San Francisco: ASP), 335

\bibitem[Merritt, Ferrarese, \& Joseph(2001)]{mer01b} Merritt, D.,
Ferrarese, L., \& Joseph, C. L. 2001, Science, 293, 1116

\bibitem[Moshir et al.(1990)]{mos90} Moshir, M., et al. 1990,
IRAS Faint Source Catalog, Version 2.0

\bibitem[Muxlow et al.(1994)]{mux94} Muxlow, T. W. B., Pedlar, A.,
Wilkinson, P. N., Axon, D. J., Sanders, E. M., \& de Bruyn, A. G.
1994, \mnras, 266, 455

\bibitem[Nagar et al.(2000)]{nag00} Nagar, N. M., Falcke, H.,
Wilson, A. S., \& Ho, L. C. 2000, \apj, 542, 186

\bibitem[Nagar et al.(2002)]{nag02} Nagar, N. M., Falcke, H., Wilson, A. S.,
\& Ulvestad, J. S. 2002, \aap, in press (astro-ph/0207176)

\bibitem[Neff \& Ulvestad(2000)]{nef00} Neff, S. G., \&
Ulvestad, J. S. 2000, \aj, 120, 670

\bibitem[Sandage \& Tammann(1981)]{san81} Sandage, A. R., \&
Tammann, G. A. 1981, A Revised Shapley-Ames Catalog of Bright
Galaxies (Washington, DC: Carnegie Inst. of Washington) 

\bibitem[Smith et al.(1998)]{smi98} Smith, H. E., Lonsdale, C. J.,
Lonsdale, C. J., \& Diamond, P. J. 1998, \apjl, 493, L17

\bibitem[Soifer et al.(1989)]{soi89} Soifer, B. T., Boehmer, L.,
Neugebauer, G., \& Sanders, D. B. 1989, \aj, 98, 766

\bibitem[Taylor et al.(1999)]{tay99} Taylor, G. B., Silver, C. S.,
Ulvestad, J. S., \& Carilli, C. L. 1999, \apj, 519, 185

\bibitem[Thompson et al.(1980)]{tho80} Thompson, A. R., Clark, B. G.,
Wade, C. M., \& Napier, P. J. 1980, \apjs, 44, 151

\bibitem[Tremaine et al.(2002)]{tre02} Tremaine, S., et al. 2002, \apj, 
574, 740

\bibitem[Tully(1988)]{tul88} Tully, R. B. 1988, Nearby Galaxies Catalogue
(Cambridge: Cambridge Univ. Press)

\bibitem[Turner, Beck, \& Ho(2000)]{tur00} Turner, J. L., Beck, S. C.,
\& Ho, P. T. P. 2000, \apjl, 532, L109

%\bibitem[Ulvestad(1982)]{ulv82} Ulvestad, J. S. 1982, \apj, 259, 96

\bibitem[Ulvestad \& Antonucci(1997)]{ulv97} Ulvestad, J. S., \&
Antonucci, R. R. J. 1997, \apj, 488, 621

\bibitem[Ulvestad \& Ho(2001)]{ulv01} Ulvestad, J. S., \& Ho, L. C. 
2001, \apj, 558, 561

\bibitem[Van Dyk et al.(1999)]{van99} Van Dyk, S. D., Lacey, C. K.,
Sramek, R. A., \& Weiler, K. W. 1999, IAUC 7322

\bibitem[van Moorsel, Kemball, \& Greisen(1996)]{van96}
van Moorsel, G., Kemball, A., \& Greisen, E. 1996, in Astronomical
Data Analysis Software and Systems V, ed. G. H. Jacoby \&
J. Barnes (San Francisco: ASP), 37

%\bibitem[Wandel(2001)]{wan01} Wandel, A. 2001, in Probing the Physics
%of Active Galactic Nuclei by Multiwavelength Monitoring,
%ASP Conf. Ser. 224, ed. B. M. Peterson, R. S. Polidan, \& R. W. Pogge
%(San Francisco: ASP), 365

\end{thebibliography}
\end{document}